# A note on Solving Parametric Polynomial Systems

Asieh Pourhaghani

Department of Mathematical Sciences, Isfahan University of Technology,
Isfahan, 84156-83111, Iran
asieh.pourhaghani@math.iut.ac.ir

**Abstract.** Lazard and Rouillier in [9], by introducing the concept of discriminant variety, have described a new and efficient algorithm for solving parametric polynomial systems. In this paper we modify this algorithm, and we show that with our improvements the output of our algorithm is always minimal and it does not need to compute the radical of ideals.

**Key words:** Parametric polynomial system, Constructible set, Discriminant variety, Gröbner basis, Elimination ideal.

## 1 Introduction

Many problems in science and engineering such as biology, chemistry, computer science, robotics, and so on, can be reduced to solving a parametric polynomial system. In 1948, Tarski [13], has published a quantifier elimination method for the elementary theory of real closed fields. However the Tarski method provides a decision method, which enables one to decide truth any sentence of the theory, but the complexity of this approach makes it unsuitable for non trivial problems. Later, Böge [7] made significant improvement to this method. We may also cite the work of Seidenberg [12], and later Cohen [2] on this subject. Then Collins introduced a completely new method called Cylindrical Algebraic Decomposition (CAD), see [3].

Weispfenning in [16], using the concept of comprehensive Gröbner bases, has presented an algorithm to describe the complex solutions of a parametric system, but his method gives no information on its real roots. Triangular decomposition (see [8,1,15,14] for example) is another approach that can be used to decompose a parametric system. Complexity results for this method can be found in [5].

Lazard and Rouillier [9], have introduced the concept of *discriminant variety*. Using a projection the parameters space is divided into two parts: the discriminant variety and its complement. The discriminant variety is a part for which the system has a non-generic behavior (see Section 2 for more details). In [10], Moroz has proved that the degree of the computed minimal discriminant variety is bounded by $D := (n-d+\ell)k^{(n-d+1)}$ and the variety can be computed within the bit complexity $\sigma^{\mathcal{O}(1)} D^{\mathcal{O}(n)}$ if the input system has $n-d$ polynomial equations, $\ell$ polynomial inequations in $n-d$ variables of degree bounded by $k$ with coefficients in a polynomial ring of $d$ parameters with rational coefficients of bit-size at most $\sigma$.

To compute two component of discriminant variety (critical and singular varieties), by the algorithms described in [9], we may need to compute the radical and primary decomposition of some ideals. Moreover, the computed discriminant variety is not always minimal. In this paper, we give some improvements of this algorithm such that to compute these two components, we do not need the computation of primary decomposition and radical of ideals. Also, we prove that, with our improvements, the computed discriminant variety is minimal (see Section 3).

In Section 2, we recall the definition of the discriminant variety from [9]. Section 3 is devoted to computing two components of the discriminant variety without using primary decomposition and radical of ideals. In section 4, we compare by an example, computing these two components, by the algorithm in [9] and our algorithm.

## 2   Discriminant Variety

In this section, we recall the definition of discriminant variety, and we state some results from [9] to compute it. For more details see [9]. Throughout this paper, we use the following notation.

**Notation 1** *Let $R = \mathbb{C}[u_1, \ldots, u_d, x_{d+1}, \ldots, x_n]$ be a polynomial ring where $\mathbb{C}$ is the field of complex number, $U = \{u_1, \ldots, u_d\}$ is the set of parameters and $X = \{x_{d+1}, \ldots, x_n\}$ is the set of variables. Let also $p_1, \ldots, p_s, f_1, \ldots, f_\ell \in R$ be some polynomials. Let $\mathcal{E} = \{p_1, \ldots, p_s\}$ is the set of equalities and $\mathcal{F} = \{f_1, \ldots, f_\ell\}$ is the set of inequalities. we define*
$\mathcal{C} = \{x \in \mathbb{C}^n \mid p_1(x) = \cdots = p_s(x) = 0, f_1(x) \neq 0, \ldots, f_\ell(x) \neq 0\}$.
*We denote by $\Pi_U : \mathbb{C}^n \longrightarrow \mathbb{C}^d$ the projection map $\Pi_U(U,X) = U$. Let $\overline{\Pi_U(\mathcal{C})}$ be the $\mathbb{Q}$-Zariski closure of $\Pi_U(\mathcal{C})$ and $\delta$ be its dimension.*

We can easily prove that $\overline{\Pi_U(\mathcal{C})} = \overline{\Pi_U(\overline{\mathcal{C}})}$.

**Definition 1.** *The affine variety $\mathcal{W} \subset \overline{\Pi_U(\mathcal{C})} \subset \mathbb{C}^d$ is called discriminant variety of $\mathcal{C}$ w.r.t. $\Pi_U$ if the followings hold:*

1. $\mathcal{W} \subsetneq \overline{\Pi_U(\mathcal{C})}$.
2. $\mathcal{W} = \overline{\Pi_U(\mathcal{C})}$ if and only if $\Pi_U^{-1} \cap \mathcal{C}$ is infinite for all $u \in \overline{\Pi_U(\mathcal{C})}$.
3. $\overline{\Pi_U(\mathcal{C})} \backslash \mathcal{W}$ is union of finite number of connected component of dimension $\delta$ (i.e. $\overline{\Pi_U(\mathcal{C})} = \bigcup_{i=1}^{k} \mathcal{U}_i$ where $\mathcal{U}_i$ is a connected component of $\overline{\Pi_U(\mathcal{C})} \backslash \mathcal{W}$ and $\dim(\mathcal{U}_i) = \delta$ for all $1 \leq i \leq k$).
4. $\overline{\Pi_U(\mathcal{C})} \backslash \mathcal{W}$ is a covering space of $\mathcal{U}_i$ for all $1 \leq i \leq k$.

From the fourth property of the above definition, we can conclude that for each connected component $\mathcal{U} \subset \overline{\Pi_U(\mathcal{C})} \backslash \mathcal{W}$, there exist a finite set of indices $\Lambda$ and disjoint connected component subsets $\{\mathcal{V}_\lambda\}_{\lambda \in \Lambda}$ of $\mathcal{C}$ such that $\Pi_U^{-1}(\mathcal{U}) \cap \mathcal{C} = \bigcup_{\lambda \in \Lambda} \mathcal{V}_\lambda$ and $\Pi_U \mid_{\mathcal{V}_\lambda} : \mathcal{V}_\lambda \to \mathcal{U}$ is a diffeomorphism for all $\lambda \in \Lambda$.

Now we introduce some other notations that use through this paper.

**Notation 2**

1. $O_\infty$ is the set of $u \in \overline{\Pi_U(\mathcal{C})}$ such that $\Pi_U^{-1}(\mathcal{U}) \cap \overline{\mathcal{C}}$ is not compact for any compact neighborhood $\mathcal{U}$ of $u$.
2. $O_{sd}$ is the projection by $\Pi_U$ of the components of dimension less than $\delta$ of $\overline{\mathcal{C}}$.
3. $O_c$ is the projection by $\Pi_U$ of the critical locus of the union of the components of dimension $\delta$ of $\overline{\mathcal{C}}$.
4. $O_\mathcal{F}$ is the projection by $\Pi_U$ of the intersection of $\mathcal{C}$ with $\prod_{i=1}^\ell f_i = 0$ (i.e. $O_\mathcal{F} = \Pi_U(\overline{\mathcal{C}} \cap \mathrm{V}(\prod_{i=1}^\ell f_i)))$.
5. $O_{sing}$ is the singular locus of $\overline{\Pi_U(\mathcal{C})}$.
6. $\mathcal{W}_x$ is the Zariski closure of $O_x$.

The affine variety $O_\infty \cup O_{sd} \cup O_c \cup O_\mathcal{F} \cup O_{sing}$ is contained in any discriminant variety. Now we show that it is the minimal discriminant variety.

**Proposition 1.** $\mathcal{W}_D = O_\infty \cup O_{sd} \cup O_c \cup O_\mathcal{F} \cup O_{sing} = \mathcal{W}_\infty \cup \mathcal{W}_{sd} \cup \mathcal{W}_c \cup \mathcal{W}_\mathcal{F} \cup \mathcal{W}_{sing}$ is the minimal discriminant variety of $\mathcal{C}$ w.r.t. $\Pi_U$.

*Proof.* See [9], Theorem 1.

Now, the main problem is how we can compute $\mathcal{W}_D$ by the above proposition. Let $\mathcal{I} \subset \mathbb{Q}[U, X]$ be an ideal such that $\mathrm{V}(\mathcal{I}) = \overline{\mathcal{C}}$. By PREPROCESSING (resp. PROPERNESSDEFECTS) algorithm in [9] we can compute $\mathcal{I}$, $\overline{\Pi_U(\mathcal{C})}$, $\delta$ and $\mathcal{W}_\mathcal{F}$ (resp. $\mathcal{W}_\infty$). For computing $\mathcal{W}_{sd}$ we may need to compute a primary decomposition of $\mathcal{I}$, however for the real examples it is empty. In the next section, we will propose some improvements of the algorithms described in [9] to compute $\mathcal{W}_c$ and $\mathcal{W}_{sing}$ for computing $\mathcal{W}_D$.

## 3 Statement of the main results

In this section, we state our main results to compute $\mathcal{W}_c$ and $\mathcal{W}_{sing}$. For this, we need the following notations.

**Notation 3** *Let $Y = [u_1, \ldots, u_d, x_{d+1}, \ldots, x_n]$ and $I \subset \mathbb{Q}[Y]$ be an ideal. Let $Y' \subset Y$ be a subset of the variables and $k \leq \#Y'$ be a positive integer. We denote by $\mathrm{Jac}_{Y'}^k(I)$ the ideal generated by minors of size $k$ of the Jacobian matrix w.r.t. $Y'$ of a system of generators of $I$.*

Since $\mathbb{Q}[U, X]$ is a Cohen-Macaulay ring, the codimension of $\mathcal{I}$ (see the previous section) is equal to $n - \delta$. Thus Jacobian criterion and definition of $\mathcal{W}_c$ and $\mathcal{W}_{sing}$ imply that

$$\mathcal{W}_c = \mathrm{V}((\mathcal{I} + \mathrm{Jac}_X^{n-\delta}(\mathcal{I})) \cap \mathbb{Q}[U])$$
$$\mathcal{W}_{sing} = \mathrm{V}((\mathcal{I} \cap \mathbb{Q}[U]) + \mathrm{Jac}_U^{d-\delta}(\mathcal{I} \cap \mathbb{Q}[U])).$$

We will show that to compute $\mathcal{W}_c$ and $\mathcal{W}_{sing}$, one can replace $\mathrm{Jac}_X^{n-\delta}(\mathcal{I})$ and $\mathrm{Jac}_U^{d-\delta}(\mathcal{I}\cap\mathbb{Q}[U])$ by $\mathrm{Jac}_X^{n-\delta}(\langle\mathcal{E}\rangle)$ and $\mathrm{Jac}_U^{d-\delta}(\langle\mathcal{E}\rangle\cap\mathbb{Q}[U])$ respectively. The computation of $\mathrm{Jac}_Y^{n-\delta}(\mathcal{I})$ is more expensive than the computation of $\mathrm{Jac}_Y^{n-\delta}(\langle\mathcal{E}\rangle)$ for any $Y\subset\{U,X\}$, since the number of component of $\mathcal{I}$ is more than the number of component of $\mathcal{E}$.

**Lemma 1.** *The ideals $\mathcal{I}+\mathrm{Jac}_X^{n-\delta}(\mathcal{I})$ and $\mathcal{I}+\mathrm{Jac}_X^{n-\delta}(\langle\mathcal{E}\rangle)$ (and $\langle\mathcal{E}\rangle+\mathrm{Jac}_X^{n-\delta}(\langle\mathcal{E}\rangle)$) have the same zeros outside $\mathrm{V}(\prod_{i=1}^{\ell}f_i)$.*

*Proof.* Let $f=\prod_{i=1}^{\ell}f_i$. Since $\mathcal{I}=\langle\mathcal{E}\rangle:f^\infty$, then $\langle\mathcal{E}\rangle\subset\mathcal{I}$ and therefore $\mathrm{V}(\mathcal{I})\subset\mathrm{V}(\langle\mathcal{E}\rangle)$. Conversely, let $g\in\mathcal{I}$. Then by definition, there exists $t\in\mathbb{N}$ such that $gf^t\in\langle\mathcal{E}\rangle$. Let $a\in\mathrm{V}(\langle\mathcal{E}\rangle)$ and $f(a)\neq 0$, so $(gf^t)(a)=0$, then $g(a)=0$. This follows that $a\in\mathrm{V}(\mathcal{I})$ and $\mathrm{V}(\langle\mathcal{E}\rangle)\backslash\mathrm{V}(f)\subset\mathrm{V}(\mathcal{I})$. Thus $\mathrm{V}(\mathcal{I})\backslash\mathrm{V}(f)=\mathrm{V}(\langle\mathcal{E}\rangle)\backslash\mathrm{V}(f)$.

Now let $\mathcal{I}=\langle g_1,\ldots,g_k\rangle$. The Jacobian matrix of the ideal $\mathcal{I}$ does not depend on the chosen generators of $\mathcal{I}$ (see [6], Corollary 16.20, page 405). Since $\langle\mathcal{E}\rangle\subset\mathcal{I}$, then $\mathrm{Jac}_X^{n-\delta}(\langle\mathcal{E}\rangle)\subset\mathrm{Jac}_X^{n-\delta}(\mathcal{I})$, so $\mathrm{V}(\mathrm{Jac}_X^{n-\delta}(\mathcal{I}))\subset\mathrm{V}(\mathrm{Jac}_X^{n-\delta}(\langle\mathcal{E}\rangle))$. Conversely by definition of $\mathcal{I}$, for every $g_i\in\{g_1,\ldots,g_k\}$ there exists a natural number $t_i$ such that $g_if^{t_i}\in\langle\mathcal{E}\rangle$. Set $t=\max\{t_i\mid 1\le i\le k\}$. Then $g_if^t\in\langle\mathcal{E}\rangle$ for any $i$, and $\mathrm{Jac}_X(f^t\mathcal{I})\subset\mathrm{Jac}_X(\langle\mathcal{E}\rangle)$. Thus

$$\mathrm{Jac}_X(f^t\mathcal{I})=\left(\frac{\partial g_if^t}{\partial x_j}\right)_{k\times d}=\left(f^t\frac{\partial g_i}{\partial x_j}+g_i\frac{\partial f^t}{\partial x_j}\right)_{k\times d}.$$

Now let $a\in\mathrm{V}(\mathcal{I})\backslash\mathrm{V}(f)$, then $g_i(a)=0$ for $1\le i\le k$, so

$$\mathrm{Jac}_X(f^t\mathcal{I})(a)=\left(f^t(a)\frac{\partial g_i(a)}{\partial x_j}\right)_{k\times d}$$
$$=f^t(a)\left(\frac{\partial g_i(a)}{\partial x_j}\right)_{k\times d}=f^t(a)\mathrm{Jac}_X(\mathcal{I})(a)$$

thus $\mathrm{V}(\mathrm{Jac}_X^{n-\delta}(\langle\mathcal{E}\rangle))\backslash\mathrm{V}(f)\subset\mathrm{V}(\mathrm{Jac}_U^{n-\delta}(\mathcal{I}))\backslash\mathrm{V}(f)$ and it ends the proof.

The computation of $(\langle\mathcal{E}\rangle+\mathrm{Jac}_X^{n-\delta}(\langle\mathcal{E}\rangle))\cap\mathbb{Q}[U]$, in practice, is usually faster than the computation of $(\mathcal{I}+\mathrm{Jac}_X^{n-\delta}(\langle\mathcal{E}\rangle))\cap\mathbb{Q}[U]$. So it would be better if we replace $\mathcal{I}$ by $\langle\mathcal{E}\rangle$ to calculate $\mathcal{W}_c$ and $\mathcal{W}_{sing}$. For instance, in [9], Example 5.1 (cuspidal manipulators) if we compute $(\langle\mathcal{E}\rangle+\mathrm{Jac}_X^{n-\delta}(\langle\mathcal{E}\rangle))\cap\mathbb{Q}[U]$ by the corresponding algorithm that we have implemented in MAPLE 12, it takes 11133 seconds, while for computing $(\mathcal{I}+\mathrm{Jac}_X^{n-\delta}(\langle\mathcal{E}\rangle))\cap\mathbb{Q}[U]$, we need 15453 seconds (the timings were conducted on a personal computer with Intel(R) Core(Tm)2 Duo CPU T5670@1.80 GHz and 1.79 GHz, 1.99 GB of RAM).

The proof of the above theorem implies the following results.

**Corollary 1.** *The ideals $(\mathcal{I}\cap\mathbb{Q}[U])+\mathrm{Jac}_U^{d-\delta}(\mathcal{I}\cap\mathbb{Q}[U])$ and $(\mathcal{I}\cap\mathbb{Q}[U])+\mathrm{Jac}_U^{d-\delta}(\langle\mathcal{E}\rangle\cap\mathbb{Q}[U])$ have the same zeros outside $\mathrm{V}(\langle\prod_{i=1}^{\ell}f_i\rangle\cap\mathbb{Q}[U])$.*

**Theorem 1.** $\mathcal{W}_\infty \cup \mathcal{W}_{sd} \cup \mathcal{W}_\mathcal{F} \cup V((\mathcal{I} + \mathrm{Jac}_X^{n-\delta}(\langle\mathcal{E}\rangle)) \cap \mathbb{Q}[U]) \cup V((\mathcal{I} \cap \mathbb{Q}[U]) + \mathrm{Jac}_U^{d-\delta}(\langle\mathcal{E}\rangle \cap \mathbb{Q}[U]))$ *is equal to* $\mathcal{W}_D$, *and therefore it is the minimal discriminant variety.*

It is worth noting that Lazrad and Rouillier in [9], have proved this theorem in the case that $\dim(((\mathcal{I} + \mathrm{Jac}_X^{n-\delta}(\mathcal{I})) \cap \mathbb{Q}[U])) < \delta$ (see [9], Proposition 3). Now, we describe CRITICAL algorithm to compute $\mathcal{W}_c$.

---

CRITICAL **algorithm**

**Input:** $\mathcal{E}$, $G$, $\delta$, $U$, $X$ where $G$ is the reduced Gröbner basis of $\mathcal{I}$ w.r.t. $\prec_{U,X}$ where $\prec_U$ and $\prec_X$ are degree reverse lexicographical ordering.
**Output:** The reduced Gröbner basis $G_c$ w.r.t. $\prec_U$ such that $\mathcal{W}_\mathcal{F} \cup \mathcal{W}_c = \mathcal{W}_\mathcal{F} \cup V(\langle G_c \rangle)$.
$G_{jac}:=$ The reduced Gröbner basis of $\langle G \rangle \cup \mathrm{Jac}_X^{n-\delta}(\langle\mathcal{E}\rangle)$ w.r.t. $\prec_{U,X}$;
$G_c := G_{jac} \cap \mathbb{Q}[U]$;
**Return** $(G_c)$;

---

We present SINGULAR algorithm to compute $\mathcal{W}_{sing}$. Note that if $\delta = d$, then $W_{sing} = \emptyset$ to optimize the computation.

---

SINGULAR **Algorithm**

**Input:** $\mathcal{E}$, $G_\Pi$, $\delta$, $U$, $X$ where $G_\Pi$ is the reduced Gröbner basis of $\mathcal{I} \cap \mathbb{Q}[U]$ w.r.t. $\prec_U$ where $\prec_U$ is degree reverse lexicographical ordering.
**Output:** The reduced Gröbner basis $G_{sing}$ w.r.t. $\prec_U$ such that $\mathcal{W}_\mathcal{F} \cup \mathcal{W}_{sing} = \mathcal{W}_\mathcal{F} \cup V(\langle G_{sing} \rangle)$.
**if** $\delta < d$ **then**
    $G_{sing}:=$ The reduced Gröbner basis of $\langle G_\Pi \rangle \cup \mathrm{Jac}_U^{d-\delta}(\langle G_\Pi \rangle)$ w.r.t. $\prec_U$;
**else**
    $G_{sing} = \{1\}$;
**end if**;
**Return** $(G_{sing})$;

---

## 4 Example

In this section by an example we compare the algorithm in [9] and our algorithm described in Section III to compute the discriminant variety. In this example we will show, whereas by the CRITICAL algorithm in [9] we need to compute radical of ideals and the output is not the minimal discriminant variety while by our algorithm we obtain the minimal discriminant variety without using radical of ideals.

Let $\mathcal{C} = \{(a, r, x, y) \in \mathbb{C}^4 \mid ax^2y + 5ay^3 - r^3 = 0,\ a - r^2 = 0,\ r > 0\}$ be a constructible set. Thus $\mathcal{E} = \{ax^2y + 5ay^3 - r^3, a - r^2\}$ is the set of equalities, $\mathcal{F} = \{r\}$ is the set of inequalities, $U = [r, a]$ are the parameters and $X = [x, y]$ are the variables. By the PREPROCESSING algorithm we get

$$\delta = 1$$
$$\mathcal{I} = \langle r^2 - a, 5ay^3 + ax^2y - ra \rangle$$
$$\overline{\Pi_U(\mathcal{I})} = \langle r^2 - a \rangle$$
$$\mathcal{W}_\mathcal{F} = ar.$$

The Output of PROPERNESSDEFECTS algorithm is

$$\mathcal{W}_\infty = r^2 - a.$$

Using SMALLDIMENSION algorithm, we have to compute primary decomposition of $\mathcal{I}$ which is $\mathcal{I} = \langle a, r^2 \rangle \cap \langle r^2 - a, x^2y + 5y^3 - r \rangle$ and therefore $\mathcal{W}_{sd} = \emptyset$.

If we run CRITICAL algorithm in [9], on this constructible set, we get

$$G_c = r^2 - a$$
$$G_{sing} = \emptyset$$
$$Property = \text{``NeedRadical''}$$

Thus, by the algorithm in [9], to calculate the discriminant variety we need to compute radical of ideals and the computed discriminant variety is not minimal while by our algorithm $\mathcal{W}_c = \emptyset$ and $\mathcal{W}_{sing} = \emptyset$ and the computed discriminant variety is minimal.

# References

1. P. Aubry, D. Lazard, M. Moreno Maza. On the theories of triangular sets. *Journal of Symbolic Computation*, 28(1–2):105–124, 1999.
2. P. J. Cohen. Decision Procedures for Real and P-Adic Fields. *Comput. and Appl. Math.*, 22(2), pages 131–151, 1969.
3. G. E. Collins. Quantifier Elimination for Real Closed Fields by Cylindrical Algebraic Decomposition. *Lecture Notes in Computer Science*, 33, pages 134–183, 1975.
4. D. Cox and J. Little and D. O'Shea. *Ideals, Varieties, and Algorithms An Introduction to Computational Algebraic Geometry and Commutative Algebra*. Springer-Verlag, 2007.
5. X. Dahan and É. Schost. Sharp estimates for triangular sets. *ISSAC'04*, ACM Press, pages 103–110, 2004.
6. D. Eisenbud. *Commutative Algebra with a View Toward Algebraic Geometry*. Springer-Verlag, 1995.
7. C. Holthusen. *Vereinfachungen fur Tarski's Entscheidungsverfahren der Elementaren Reelen Algebra*. Diplomarbeit University of Heidelberg, 1974.
8. M. Kalkbrener. Prime decompositions of radicals in polynomial rings. *Journal of Symbolic Computation*, 18(4):365–372, 1994.

9. D. Lazard and F. Rouillier. Solving Parametric Polynomial System. *Journal of Symbolic Computation*, 42(6), pages 636–667, 2007.
10. G. Moroz. Complexity of the Resolution of Parametric Systems of Polynomial Equations and Inequations. *ISSAC'06*, ACM Press, pages 246–253, 2006.
11. E. Schost. Computing Parametric Geometric Resolutions. *Appl. Algebra Engrg. Comm. Comput.*, 13(5), pages 349–393, 2003.
12. A. Seidenberg. A New Decision Method for Elementary Algebra. *Ann. Math.* 60(2), pages 365-374, 1954.
13. A. Tarski *A Decision Method for Elementary Algebra and Geometry.* Univ. of Calif. Press, Berkely, 1951.
14. D. Wang. *Elimination Methods.* Springer-Verlag, 2001.
15. D. Wang. Computing triangular systems and regular systems. *Journal of Symbolic Computation*, 30(2), pages 221–236, 2000.
16. V. Weispfenning. Comprehensive Gröbner Bases. *Journal of Symbolic Computation*, 14(1), pages 1–29, 1992.